\documentclass[prb,showpacs]{revtex4}
\usepackage{amsmath}
\usepackage{amssymb}
\usepackage{graphicx}
\usepackage{bm}
\usepackage{dcolumn}

\begin{document}

\title{Surface state atoms and their contribution to the surface tension of
quantum liquids}
\author{A. D. Grigoriev}
\affiliation{Samara State University, Samara, Russia}
\author{A. M. Dyugaev}
\altaffiliation[Visiting address: \ ]{Max Planck Institute for the Physics of Complex Systems, Dresden D-01187
Germany}
\author{P. D. Grigoriev}
\email{grigorev@itp.ac.ru}
\affiliation{L. D. Landau Institute for Theoretical Physics, Chernogolovka, Russia}

\begin{abstract}
We investigate the new type of excitations on the surface of liquid helium.
These excitations, called surfons, appear because helium atoms have discrete
energy level at the liquid surface, being attracted to the surface by the
van der Waals force and repulsed at a hard-core interatomic distance. The
concentration of the surfons increases with temperature. The surfons
propagate along the surface and form a two-dimensional gas. Basing on the
simple model of the surfon microscopic structure, we estimate the surfon
activation energy and effective mass for both helium isotopes. We also
calculate the contribution of the surfons to the temperature dependence of
the surface tension. This contribution explains the great and long-standing
discrepancy between theory and experiment on this temperature dependence in
both helium isotopes. The achieved agreement between our theory and
experiment is extremely high. The comparison with experiment allows to
extract the surfon activation energy and effective mass. The values of these
surfon microscopic parameters are in a reasonable agreement with the
calculated from the proposed simple model of surfon structure.
\end{abstract}

\pacs{73.20.Dx,67.55.S}
\keywords{liquid surface, surface tension, surfon}
\maketitle

\section{Introduction}

The microscopic description of the surface of liquids touches various fields
of natural science. This problem is not simple, and even the calculation of
the surface tension coefficient has many difficulties.\cite{SurfTensionBook}
At low temperature the quantum nature of the surface excitations becomes
important. At very low temperature only few types of surface excitations
with the energy less or of the order of temperature are relevant for the
problem. This fact greatly simplifies the description of liquid surface in
the low-temperature limit. An accurate calculation of the absolute value of
the surface tension coefficient $\alpha $ remains a challenge, but the
calculation of its temperature dependence $\alpha \left( T\right) $ is much
simpler. Among all liquids, the low-temperature limit is reached only in
helium, and the surface of liquid helium has been studied experimentally in
great detail.\cite{EdwardsReview1978} Therefore, we apply our analysis
mainly to the liquid helium. The experimental values of the surface tension
of liquid $^{4}$He and $^{3}$He at zero temperature are\cite%
{SurTensExp4,SurTensExp3} $\alpha _{He4}\left( T=0\right) \equiv \alpha
_{04}=0.3544\,$dyn/cm and $\alpha _{He3}\left( T=0\right) \equiv \alpha
_{03}=0.155$\thinspace dyn/cm. Taking these values as a reference point, one
may calculate the deviation $\Delta \alpha \left( T\right) \equiv $ $\alpha
_{0}-\alpha \left( T\right) $ as the sum of the contributions from all types
of surface excitations to the surface free energy per unit square. At low
temperature the concentration of these excitations is low, and the
interaction between the excitations can be neglected. Then the surface
excitations form an ideal two-dimensional gas of particles with the
dispersion determined by the nature of these excitations. The free energy of
this gas is well known\cite{LL5} [see Eq. (\ref{OmSigma}) below].

At sufficiently low temperatures, the only considered type of surface
excitations are the quanta of surface waves, called ripplons. The ripplons
lead to the temperature dependence of the surface tension of a liquid given
by the Atkins formula \cite{Atkins}:%
\begin{equation}
\Delta \alpha _{R}\left( T\right) =AT^{7/3},  \label{Atkins}
\end{equation}%
where the coefficient $A=6.8$ mdyn/cm$\cdot $K$^{7/3}$ for $^{4}$He.
However, this estimate of the temperature dependence of the surface tension
is much weaker than the measured one.\cite{SurTensExp4} Moreover, Eq. (\ref%
{Atkins}) applies only for $^{4}$He in the superfluid state, because above
the $\lambda $-point, $T_{\lambda }=2.17$K, the short-wavelength ripplons
are damped by the liquid viscosity. The damping of the high-energy
short-wavelength ripplons must lead to the strong modification of the
dependence (\ref{Atkins}) above the $\lambda $-point. On contrary, experiment%
\cite{SurTensExp4} gives only very weak change in the temperature dependence
of the surface tension at the $\lambda $-point, suggesting that the ripplon
contribution is not the main one. In liquid $^{3}$He the viscosity is much
higher than in $^{4}$He, and only the very long-wavelength ripplons with
energy $\hbar \omega _{k}\ll T$ survive at $T<1K$. These long-wavelength
ripplons give negligible contribution to the free energy because of the
small number of quantum states, which is proportional to the phase volume.
Hence, the theory predicts a very weak dependence $\alpha \left( T\right) $
for $^{3}$He in the whole temperature interval, which strongly contradicts
the experimental observations\cite{SurTensExp3}. This discrepancy between
the theory and experiment on the liquid helium surface tension remained a
puzzle for several decades, until the new type of surface excitations has
been proposed.\cite{SurStates} This new type of excitations, called the
surface level atoms (SLA) or \textit{surfons}, allowed to explain the
temperature dependence of the surface tension and to reach the very high
agreement between theory and experiment.\cite{SurStates} The surfons
resemble the states of $^{3}$He atoms on the surface of liquid $^{4}$He in
the $^{3}$He-$^{4}$He mixtures \cite{AndreevLevels} and the states of $^{3}$%
He or $^{4}$He atoms on the surface of liquid hydrogen.\cite{Hydrogen1} The
surface states in the $^{3}$He-$^{4}$He mixtures, called the Andreev states,
were also introduced to explain the temperature dependence of the surface
tension of these mixtures. The main difference between the Andreev states
and the surfons is that the latter exist even in the pure He isotopes.

A thorough microscopic description of this new type of excitations is a
rather complicated many-particle problem. In Ref. \cite{SurStates} the
surfons were considered phenomenologically as the quantum states of helium
atoms localized at the liquid helium surface. The surfons may propagate in
the surface plane and have the quadratic dispersion%
\begin{equation}
\varepsilon (k)=E_{s}+k^{2}/2M^{\ast },  \label{Disp}
\end{equation}%
where $k$ is the 2D momentum of surfons along the surface. Both the surfon
energy $E_{s}$ and their effective mass $M^{\ast }$ depend on the He isotope
$^{3}$He or $^{4}$He. The surfon energies $E_{s}$ are intermediate between
the energy of a He atom in vacuum $E_{vac}^{He}$\ and the chemical potential
$\mu $ of this atom inside the liquid. If one takes the energies of He atoms
in vacuum as the reference point, $E_{vac}^{He}=0$, the chemical potentials
of liquid $^{4}$He and $^{3}$He at $T\rightarrow 0$ are
\begin{equation}
\mu _{0}^{He4}=-7.17K,~\mbox{and}~\mu _{0}^{He3}=-2.5K.  \label{mu0}
\end{equation}%
The creation of a surfon is a thermal activation process with the activation
energy
\begin{equation}
\Delta \left( T\right) =E_{s}-\mu \left( T\right) .  \label{Delta}
\end{equation}%
Therefore, at low enough temperature $T\ll \Delta $ the concentration of
surfons is exponentially small. Fitting the experimental data on the
temperature dependence of the surface tension, below we obtain the values of
this activation energy for $^{4}$He and $^{3}$He:
\begin{equation}
\Delta ^{He4}\approx 2.67K,~\Delta ^{He3}\approx 0.25K.  \label{D}
\end{equation}%
The corresponding energies of surfons are
\begin{equation}
E_{s}^{He4}\approx -4.5K\mbox{ and }E_{s}^{He3}\approx -2.25K,
\label{SLAEnergy}
\end{equation}%
The effective masses, obtained from this fitting, are
\begin{equation}
M_{4}^{\ast }\approx 2.65M_{4}^{0},~\mbox{and}~M_{3}^{\ast }\approx
2.25M_{3}^{0},  \label{M}
\end{equation}%
where $M_{4}^{0}$ and $M_{3}^{0}$\ are the masses of free $^{4}$He and $^{3}$%
He atoms.

The role of surfons may be crucial for various properties of electrons on
the liquid helium surface (see Refs. \cite{Shikin,Edelman,Monarkha} for
reviews of this area). Electron scattering on surfons reduces the mobility
of the electrons.\cite{MobilityLett} This scattering also affects the
electron transitions between the bound states on the surface, leading to the
shift and the broadening of the transition line.\cite{Edelman,WeJETPLett2008}
This is particularly important for the physical realization of quantum bits
and quantum computing with the use of electrons on the liquid helium surface.%
\cite{DykmanQC}

In the present paper we microscopically substantiate the existence of
surfons and study their properties. We estimate the energy (\ref{SLAEnergy})
ant the effective mass (\ref{M}) of surfons from the microscopic
considerations. We show that taking into account the temperature dependence
of the chemical potential of liquid helium allows to reach the agreement
between theory and experiment on the dependence $\Delta \alpha \left(
T\right) $ up to the accuracy of the existing experiments. We propose the
theoretical curve of this dependence in the entire temperature interval from
zero to the boiling temperature $T_{c}$ of liquid helium. This theoretical
prediction explains the existing puzzles in the temperature dependence of
the surface tension of liquid helium and other liquids, and stimulates
further experiments.

\section{Estimation of the surfon activation energy and effective mass}

The goal of this section is to prove the existence of the surfons from the
quantum-mechanical calculation and to give a rough estimate of its
activation energy and effective mass. In the next section we find these
values phenomenologically with higher accuracy from the comparison with
experiment on the temperature dependence of the surface tension $\Delta
\alpha \left( T\right) $. In this section we use the simplified interaction
between He atoms, which includes only the Hartree term of the van der Waals
forces and neglects the exchange interaction. Then we solve the one-particle
Schr{\"o}dinger equation for the surface atom in the potential, formed by
other atoms, taking into account possible formation of a dimple (or polaron)
under the atom on the surface.

\subsection{The surfon activation energy levels on the flat surface}

The interaction potential between two helium atoms can be described by the
Lennard-Jones (LJ) potential\cite{LJ}
\begin{equation}
V_{LJ}\left( r\right) =4\epsilon _{0}\left[ \left( \sigma _{0}/r\right)
^{12}-\left( \sigma _{0}/r\right) ^{6}\right] ,  \label{LJ}
\end{equation}%
where the generally accepted coefficients
\begin{equation*}
\epsilon _{0}=10.22K~\mbox{and}~\sigma _{0}=2.556\mathring{A}
\end{equation*}%
are the same for both He isotopes and obtained by fitting the He atom
scattering experiments. This potential includes the van-der-Waals attraction
between atoms at long distance and the hard-core repulsion at short
distance. More complicated He-He interatomic potentials have been proposed
by various authors,\cite{HeHeInt} but for our estimates the accuracy of the
potential (\ref{LJ}) is safficient. The potential energy of an atom above
the liquid can be calculated by the summation of the interatomic potential
over all atoms in the liquid. This approximation neglects the many-particle
effects and the back influence of an atom above the surface on the bulk
liquid, studied later. The integration of the Lennard-Jones potential (\ref%
{LJ}) over the uniform liquid in the half-space $z<0$ gives the following
potential energy of an atom above the surface of liquid:%
\begin{equation}
V\left( z\right) =\frac{4\pi \epsilon _{0}\sigma _{0}^{3}n_{b}}{3}\left[
\frac{1}{15}\left( \frac{\sigma _{0}}{z}\right) ^{9}-\frac{1}{2}\left( \frac{%
\sigma _{0}}{z}\right) ^{3}\right] .  \label{Vz}
\end{equation}%
Here $z$ is the distance of the atom from the surface, $n_{b}$ is the number
atom density in the bulk liquid. This bulk density in liquid $^{4}$He at $%
T\rightarrow 0$ is $n_{He4}=0.02186\mathring{A}^{-3}$, and in $^{3}$He $%
n_{He3}=0.0164\mathring{A}^{-3}$. The potential (\ref{Vz}) is attractive at
long distance $z>z_{\min }=\sigma _{0}\left( 2/5\right) ^{1/6}\approx 2.194%
\mathring{A}$ and repulsive at $z<z_{\min }$. The substitution of the
numbers to Eq. (\ref{Vz}) gives the following potentials of He atoms above $%
^{4}$He and $^{3}$He surfaces:%
\begin{eqnarray}
V^{He4}\left( z\right) &=&\left( \frac{4852}{z^{9}}-\frac{130.5}{z^{3}}%
\right) \left[ K\right] ,  \label{VHe4} \\
V^{He3}\left( z\right) &=&\left( \frac{3640}{z^{9}}-\frac{97.9}{z^{3}}%
\right) \left[ K\right] .  \label{VHe3}
\end{eqnarray}%
The numerical solution of the one-dimensional Schr{\" o}dinger equation for
a He atom in these potentials \textbf{proves the existence} of discrete
energy levels for both He isotopes, and gives the following estimates for
the energy of this bound state in the zeroth approximation:
\begin{equation}
E_{s0}^{He4}\approx -1.24K,~\mbox{and}~E_{s0}^{He3}\approx -0.342K.
\label{SLAEnergy0}
\end{equation}%
These calculated values are higher than the values (\ref{SLAEnergy})
obtained from the surface tension data fit, which is not surprising. First,
the above model takes the liquid surface to be rigid and flat, while in fact
this surface is very soft and subjected to deformation, which reduces the
bound state energy (\ref{SLAEnergy0}). Second, the distribution of He atoms
in the liquid is not uniform and is affected by an atom on the surface
level. One expects the increase of liquid density around the surfons.\
Third, the identity of the atom on the surface level with the atoms in the
bulk liquid leads to the exchange energy correction and other many-body
effects.\cite{Comment1}

The similar problem appears in the calculation of the energy of Andreev
levels,\cite{AndreevLevels} i.e. of the bound states of $^{3}$He atoms on
the surface of liquid $^{4}$He. The solution of 1D Schr{\" o}dinger equation
with the potential (\ref{Vz}) gives the energy of bound state $E_{A}\approx
-0.8K,$ while the value of Andreev energy levels, as obtained from the
surface tension experiments, is about $E_{A}^{\exp }\approx -5K$. Various
approaches have been developed to calculate the Andreev energy levels more
accurately,\cite{EdwardsReview1978} but the substantial disagreement between
theoretical and experimental results still exists.

\subsection{Formation of a dimple and its influence on the surfon activation
energy and effective mass.}

In this section we study the back influence of the surfons on the bulk
liquid. In the first approximation, one must consider the formation of a
dimple under the surfon, similar to the dimple under an electron\cite%
{Shikin,Edelman,Monarkha} or negative ion\cite{NegIons} on the liquid helium
surface. In the next approximation one also needs to account for the
increase of liquid density in the vicinity of surfon. This adjustment of the
liquid to the appearance of the surfon is similar to the polarons around
electrons in solids.\cite{Mahan} The surface deformation is a polaron of
ripplons, while the increase of the liquid density under the surfon is a
polaron of bulk phonons. This composite polaron propagates with\ the surfon
along the liquid surface, reducing the surfon activation energy and
increasing its effective mass. Since the relative variation of the liquid
density around the surfon is small, the correction to the surfon activation
energy from the density polaron is also small. On contrary, the surface
deformation around the surfon may considerably change the van der Waals
potential at the surface, and this effect must be taken into account.

For the estimate of the correction to the surfon energy and its effective
mass due to the dimple on the liquid surface, we apply the methods similar
to those in the theory of electrons\cite{Shikin,Edelman,Monarkha} and
negative ions\cite{NegIons} on the liquid helium surface, replacing the
electrostatic attraction by the van-der Waals forces. The formation of a
dimple on the surface under the surfon gains the van-der Waals energy $E_{W}$
but costs the surface tension energy $E_{surf}$ and the quantum kinetic
energy $E_{kin}$ of the surfon due to the additional confinement of its wave
function near the dimple. All these contributions depend on the shape of the
dimple and determine this shape. Their self-consistent calculation requires
the solution of the axially-symmetric 3D Schr{\" o}dinger equation for He
atom above the dimple. We do not perform the complicated calculation of the
dimple profile, because the model of a classical static surface deformation
itself is rather rough. Instead, we estimate the energy gain due to the
formation of the dimple taking its shape $\Delta z\left( x,y\right) =\xi
\left( \mathbf{\rho \equiv }\sqrt{x^{2}+y^{2}}\right) $ to be a spherical
cap of radius $R$ and depth $h$:
\begin{equation}
\xi \left( \mathbf{\rho }\right) =\left\{
\begin{array}{cc}
R-h-\sqrt{R^{2}-\mathbf{\rho }^{2}}, & ~\mathbf{\rho }<\sqrt{R^{2}-\left(
R-h\right) ^{2}} \\
0, & ~\mathbf{\rho }\geq \sqrt{R^{2}-\left( R-h\right) ^{2}}%
\end{array}%
\right.  \label{xi1}
\end{equation}%
The loss in the surface tension energy $E_{surf}$ of the shallow dimple of
depth $h\ll R$ is approximately proportional to $h^{2}$ and depends weakly
on the shape of the dimple. Thus, for the spherical dimple (\ref{xi1})
\begin{equation*}
E_{surf}=\pi h^{2}\sigma ,
\end{equation*}%
while for the much smoother Gaussian shape $\xi \left( \mathbf{\rho }\right)
=-h\exp \left( -\mathbf{\rho }^{2}/2R^{2}\right) $ with the same depth and
curvature in the center, the surface energy loss reduces by one half: $%
E_{surf}^{Gaus}=\pi h^{2}\sigma /2$. The van der Waals attraction is a very
short-range one. Therefore, the optimal dimple does not extend far from the
surfon. The optimal dimple radius is determined from the competition between
the van der Waals attraction $E_{W}$ and the kinetic energy $E_{kin}$. This
competition also happens in the bulk liquid, and leads to the mean
inter-particle distance $d=n_{b}^{-1/3}=3.58\mathring{A}$ for $^{4}$He and $%
d=3.94\mathring{A}$ for $^{3}$He. Below, we take the dimple curvature radius
$R=d$ in our estimate of the SLA energy correction. The depth of the dimple
is determined from the competition between the surface tension energy $%
E_{surf}$ and the van der Waals energy $E_{W}$.

The van der Waals energy gain from the dimple (\ref{xi1}) can be estimated
by the integration of the LJ potential (\ref{LJ}) over the region $0<z<h$
\begin{equation}
E_{W}=\int d^{3}\mathbf{r}n_{b}\left( \mathbf{r}\right) \int \Psi
_{s}^{2}\left( \mathbf{r}^{\prime }\right) d^{3}\mathbf{r}^{\prime
}V_{LJ}\left( \mathbf{r-r}^{\prime }\right) .  \label{Ew}
\end{equation}%
The kinetic energy $E_{kin}<\left\vert E_{W}\right\vert $ already implicitly
enters Eq. (\ref{Ew}) through the surfon wave function $\Psi _{s}\left(
\mathbf{r}\right) $ and the minimal dimple radius $d$. Its additional
contribution reduces the energy gain (\ref{Ew}) by the factor $\sim 1$.
However, we neglect this correction because we did not account for the
similar factor in the surface tension energy loss. We also reduce the van
der Waals energy gain by similar factor by taking the wave function in the
integral (\ref{Ew}) in the form of $\delta $-function: $\Psi _{s}^{2}\left(
\mathbf{r}^{\prime }\right) =\delta ^{3}\left( \mathbf{r}^{\prime }-\mathbf{r%
}_{0}\right) $, \ where $\mathbf{r}_{0}$ is the position of the center of
surfon. Thus, we have

\begin{equation*}
E_{W}+E_{kin}\approx n_{b}\int_{0}^{h}dz\int_{\rho _{0}\left( z\right)
}^{\infty }\pi d\rho ^{2}V_{LJ}\left( r\right) ,
\end{equation*}%
where $\rho _{0}^{2}\left( z\right) =R^{2}-\left( R-h+z\right) ^{2}\approx
2R\left( h-z\right) $ and $r^{2}=\left( R-h+z\right) ^{2}+\rho ^{2}$.
Integration gives
\begin{equation*}
\frac{E_{W}+E_{kin}}{4\pi \epsilon _{0}\sigma _{0}^{2}n_{b}}\approx h\left[
\frac{1}{5}\left( \frac{\sigma _{0}^{2}}{R^{2}}\right) ^{5}-\frac{1}{2}%
\left( \frac{\sigma _{0}^{2}}{R^{2}}\right) ^{2}\right] .
\end{equation*}%
For $^{4}$He this gives $E_{W}\approx -2.3h$ and for $^{3}$He this gives $%
E_{W}\approx -1.2h$, where the dimple depth $h$ is in $\mathring{A}$ and $%
E_{W}$ is in $K$. Minimization of
\begin{equation*}
\Delta E_{s}\approx E_{W}+E_{kin}+E_{surf}
\end{equation*}%
gives for $^{4}$He the optimal dimple depth and energy gain%
\begin{equation}
h_{4}\approx 1\mathring{A}~\mbox{and}~\Delta E_{s}^{He4}\approx -1.2K
\label{DE4}
\end{equation}%
and for $^{3}$He%
\begin{equation}
h_{3}\approx 1.2\mathring{A}~\mbox{and}~\Delta E_{s}^{He3}\approx -0.73K.
\label{DE3}
\end{equation}%
The surfon effective mass increases due to the dimple. This increase can be
roughly estimated assuming that during the surfon motion on the interatomic
distance $d$ at least all adjacent atoms must also move by the distance of
the order of the dimple depth. Then, the surfon effective mass
\begin{equation}
m^{\ast }\approx m_{He}\left( 1+n_{adj}h/d\right) .  \label{m*}
\end{equation}%
The number of adjacent atoms $n_{adj}\sim 6$, and we get
\begin{equation}
M_{4}^{\ast }\approx 3M_{4}^{0},~\mbox{and}~M_{3}^{\ast }\approx 2M_{3}^{0}.
\label{m*n}
\end{equation}

The above rough calculation is aimed to show that the surface deformation
under the surfon is essential both for the surfon activation energy. \ For
more accurate study of the surfon parameters, in the next section we compare
with experiment the thermodynamical properties of liquid surface, calculated
using the above model of surfons.

\section{Temperature dependence of liquid helium surface tension.}

The contribution $\Delta \alpha _{S}(T)$ of the 2D surfon gas to the
temperature dependence of the surface tension is\cite{LL5}
\begin{equation}
\Delta \alpha _{S}(T)=\pm gT\int \ln \left[ 1\pm \exp \left( \frac{\mu
-\varepsilon (k)}{T}\right) \right] \frac{d^{2}k}{(2\pi \hbar )^{2}}.
\label{OmSigma}
\end{equation}%
This is just a contribution of the 2D gas of noninteracting particles with
dispersion $\varepsilon (k)$ to the free energy of unit surface area.\cite%
{LL5} Here $\mu =\mu (T)$ is the temperature-dependent chemical potential of
liquid helium, $\varepsilon (k)$ is the dispersion of surfons, $g$ is the
spin degeneracy ($g=1$ for $^{4}$He and $g=2$ for $^{3}$He), and the sign "$%
\pm $" in (\ref{OmSigma}) is "$-$" for bosons and "$+$" for fermions. The
surfons above $^{4}$He are the bosons with dispersion (\ref{Disp}). After
introducing the new variable $z\equiv \exp \left[ -k^{2}/2M^{\ast }T\right] $%
, their contribution to the surface tension becomes
\begin{equation}
\Delta \alpha _{4}(T)=\frac{T^{2}M_{4}^{\ast }}{2\pi \hbar ^{2}}%
\int\limits_{0}^{1}\frac{\ln \left[ 1-z\,\exp \left( -\Delta _{4}/T\right) %
\right] }{z}dz,  \label{OmBoson}
\end{equation}%
where
\begin{equation}
\Delta _{4}\equiv \Delta _{4}\left( T\right) =E_{s}^{He4}-\mu _{4}\left(
T\right) .  \label{DeltaT4}
\end{equation}%
Above $^{3}$He the surfons are fermions, and from (\ref{OmSigma}) we obtain
after the integration by parts%
\begin{equation}
\Delta \alpha _{3}(T)=-\frac{M_{3}^{\ast }T^{2}}{\pi \hbar ^{2}}%
\int\limits_{0}^{1}\frac{\ln \left[ 1+z\,\exp \left( -\Delta _{3}/T\right) %
\right] }{z}dz,  \label{OmFermion}
\end{equation}%
where
\begin{equation}
\Delta _{3}\equiv \Delta _{3}\left( T\right) =E_{s}^{He3}-\mu _{3}\left(
T\right) .  \label{DeltaT3}
\end{equation}%
From (\ref{OmBoson}),(\ref{OmFermion}) we see that the contribution of
surfons to $\alpha _{3,4}(T)$ depends exponentially on temperature at low $%
T:\Delta \alpha \propto T^{2}\exp \left( -\Delta /T\right) $.

\begin{figure}[tbh]
\includegraphics{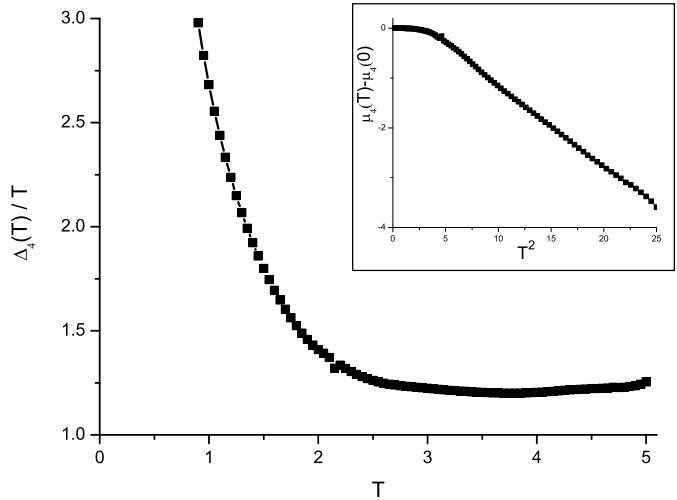}
\caption{The temperature dependence of surfon gap $\Delta _{4}\left(
T\right) $ for liquid $^{4}$He divided by temperature, as it enters Eq. (%
\protect\ref{OmBoson}). This dependence is extracted from the experimental
temperature dependence of chemical potential $\protect\mu _{4}(T)-\protect%
\mu _{4}(0)$ of liquid $^{4}$He, shown in the insert figure and obtained
from the data in Ref. \protect\cite{HeDataBook}. The graph shows that $%
\Delta \left( T\right) /T$ is almost independent of temperature in the wide
range $T_{\protect\lambda }<T\lesssim 4K$. Such a dependence of $\Delta
\left( T\right) $, being substituted in to Eq. (\protect\ref{OmBoson}),
explains the quadratic temperature dependence of the surface tension at $%
T>T_{\protect\lambda }$. }
\label{Fig1}
\end{figure}

To calculate the integrals in (\ref{OmBoson}),(\ref{OmFermion}) one need to
know the temperature dependence of the chemical potential $\mu \left(
T\right) $, which enters Eqs. (\ref{OmBoson}) and (\ref{OmFermion}) via the
surfon energy gap $\Delta $. In Ref. \cite{SurStates} this dependence has
been considered only for $^{3}$He and only qualitatively using the
approximate formula (20) of Ref. \cite{SurStates}. For quantitative analysis
in the wider temperature range we take the dependence $\mu \left( T\right) $
from the experiment. We take the exact formula for the chemical potential
\begin{equation}
\mu \left( T\right) =\frac{F_{V}\left( T\right) +P\left( T\right) }{%
n_{L}\left( T\right) }.  \label{mu}
\end{equation}%
Here the pressure $P\left( T\right) $ is equal to the saturated vapor
pressure, and$\ $the temperature-dependent particle density $n_{L}$ in
liquid He is determined from the mass density. The free energy per unit
volume is
\begin{equation}
F_{V}\left( T\right) =\int_{0}^{T}C_{V}\left( T\right) dT-TS_{V}\left(
T\right) ,  \label{Fv}
\end{equation}%
where the heat capacity $C_{V}$ and the entropy $S_{V}$ at constant volume
are also taken from experiment. For $^{4}$He the data on the temperature
dependence of the quantities $C_{V},S_{V},P,n_{L}$ entering Eqs. (\ref{mu}),(%
\ref{Fv}) are taken from Ref. \cite{HeDataBook}. For $^{3}$He the data on $%
C_{V}\left( T\right) $ and $S_{V}\left( T\right) $ are taken from Ref. \cite%
{MuTExp}, and the data on saturated pressure are taken from Ref. \cite{PHe3}%
. The data on the temperature dependence of the chemical potential and of
the surfon activation energy are plotted in Figs. \ref{Fig1} and \ref{FigMu3}%
. Note that for both helium isotopes the experimentally measured temperature
dependence of the chemical potential is in a high agreement with the
predictions of the theory of quantum nondegenerate liquids, developed in
Refs. \cite{DyugJLTP90,DyugSSR90}.
\begin{figure}[tbh]
\includegraphics[width=0.49\textwidth]{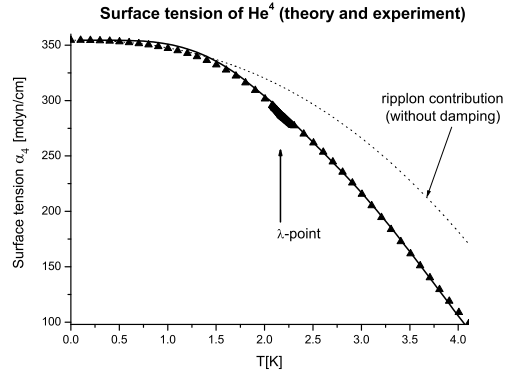}
\caption{The calculated surfon contribution to the temperature dependence of
the surface tension of $^{4}$He (solid line) compared with the experimental
data from Ref. \protect\cite{SurTensExp4} (black triangulares). At $T>T_{%
\protect\lambda }$ the agreement is so high that in the linear scale the
deviation cannot be detected by eye. In this temperature region the ripplons
are damped by viscosity, and their contribution to the surface tension $%
\Delta \protect\alpha _{4}\left( T\right) $ is negligible. At low
temperature, the concentration of surfons is exponentially small due to the
activation energy $\Delta ^{He4}\approx 2.7K$, and the main contribution to $%
\Delta \protect\alpha _{4}\left( T\right) $ comes from the gapless ripplons.
The dotted line represents Eq. (\protect\ref{Atkins}), which gives the
ripplon contribution without viscosity and is valid only deep inside the
superfluid phase. This line agrees with the experimental data only at $%
T<1.5K $.}
\label{Fig2}
\end{figure}
\begin{figure}[tbh]
\includegraphics[width=0.49\textwidth]{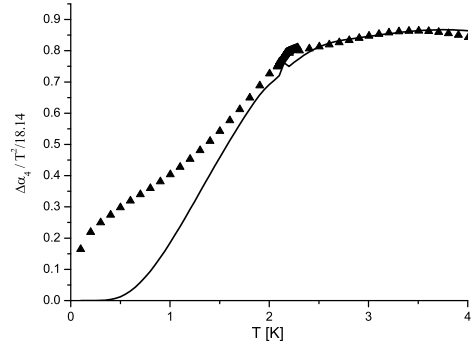}
\caption{The temperature dependence of the surface tension of $^{4}$He
divided by $T^{2}$. This graph shows the comparison between our theory and
experiment in the coordinates where the difference can be visually observed.
The low-temperature deviation at $T<T_{\protect\lambda }$ is due to the
ripplon contribution, while the deviation at $T>4K$ is due to the proximity
to the boiling point. The ripplon contribution at $T<T_{\protect\lambda }$,
extracted from this graph, is shown in Fig. \protect\ref{FigRipplons}}
\label{Fig3}
\end{figure}
\begin{figure}[tbh]
\includegraphics[width=0.49\textwidth]{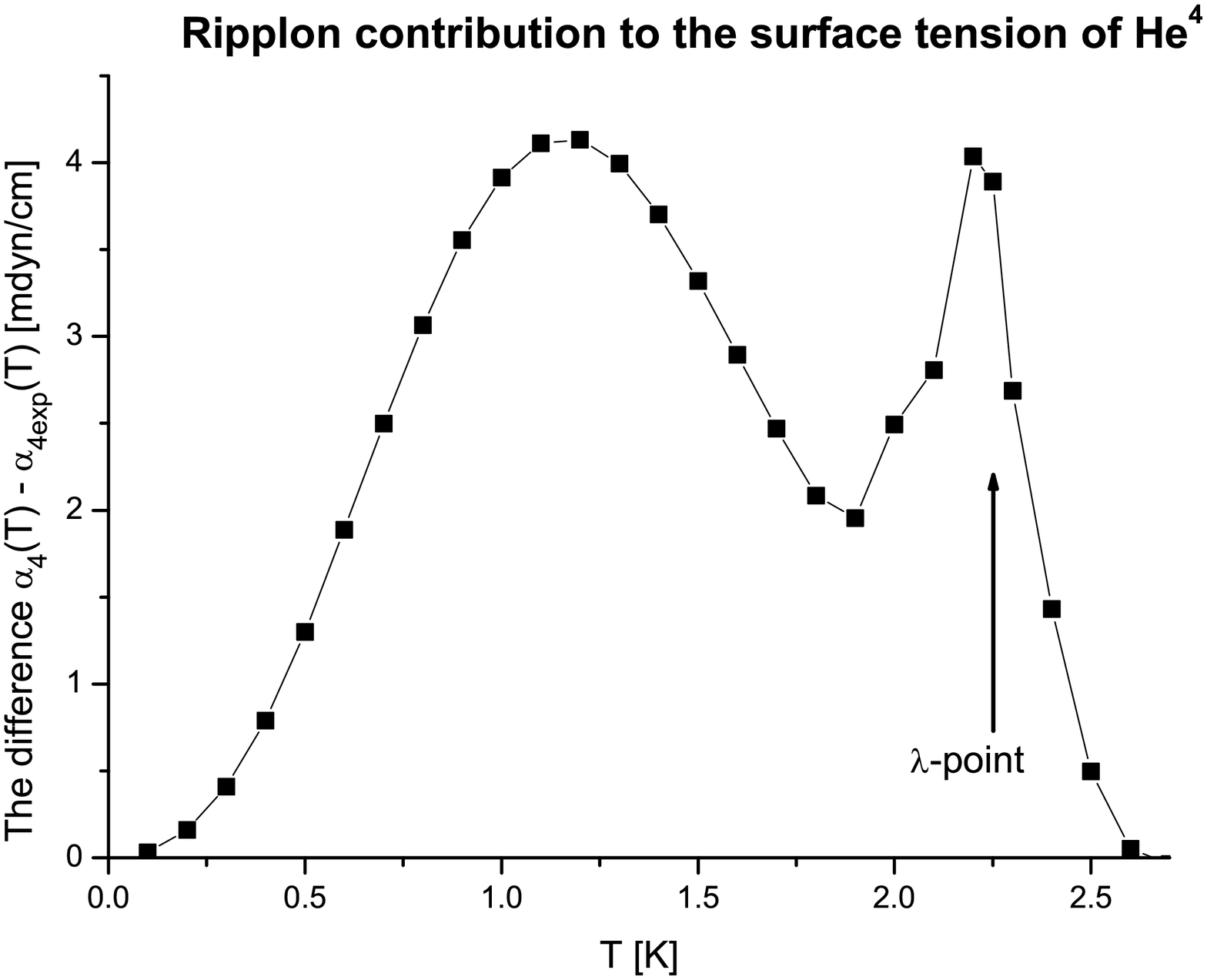}
\caption{The difference between the surfon contribution $\protect\alpha %
_{4}\left( T\right) $ to the surface tension of $^{4}$He, given by Eq. (%
\protect\ref{OmBoson}), and the experimental values $\protect\alpha _{4\exp
}\left( T\right) $. \ The maximum at $T\approx 1.1K$ is due to the ripplon
contribution. The second maximum at $T\approx 2.17K$ corresponds to phase
transition at the $\protect\lambda $-point of $^{4}$He.}
\label{FigRipplons}
\end{figure}

In the superfluid $^{4}$He at $T<T_{\lambda }=2.17K$ the chemical potential $%
\mu \left( T\right) $ depends weakly on temperature. At $T>T_{\lambda }$
this dependence is quadratic (see insert of Fig. \ref{Fig1}) in accordance
with the theoretical prediction\cite{DyugJLTP90}
\begin{equation}
\mu _{4}\left( T\right) =\mu _{4}\left( 0\right) +\varepsilon
_{4}^{0}-T^{2}/T_{4}^{0},  \label{mu4}
\end{equation}%
where $\varepsilon _{4}^{0}=0.55K$ and $T_{4}^{0}=5.59K$.\ The function $%
\Delta _{4}\left( T\right) /T$, which determines the temperature dependence
of the surface tension $\Delta \alpha _{4}(T)$ in (\ref{OmBoson}) and is
given by Eq. (\ref{DeltaT4}), depends weakly on temperature in the interval $%
T_{\lambda }<T<4.5K$ (see Fig. \ref{Fig1}). The minimum of the function $%
\Delta \left( T\right) /T$ occurs at $T\approx 3.5K$, but in the entire
range $T_{\lambda }<T<4.5K$ one can take $\Delta \left( T\right) /T=const$
with the accuracy $\sim 4\%$. Then the integrand in Eq. (\ref{OmBoson}) does
not depend on temperature, which results in the quadratic temperature
dependence of the surface tension of liquid $^{4}$He with the same accuracy $%
\sim 4\%$ (see Figs. \ref{Fig2},\ref{Fig3}). This explains the long-standing
puzzle of the experimentally observed quadratic temperature dependence of
liquid $^{4}$He at $T>T_{\lambda }$. The values $\Delta _{4}=2.67K$ and $%
M_{4}^{\ast }=2.65M_{4}^{0}$ give the best fit to the experimental points.
The ripplon contribution to the temperature dependence of the surface
tension $\Delta \alpha _{4}(T)$ is small and has maximum at $T\approx 1.1K$
(see Fig. \ref{FigRipplons}), where it makes only $\sim 2\%$ of $\alpha
_{4}\left( T\right) $ (see Figs. \ref{Fig2},\ref{Fig3}). In Ref. \cite%
{SurStates} it was shown that at $T>T_{\lambda }$ the thermal ripplons with
energy $\hbar \omega _{k}\sim T$, which give the main contribution to $%
\Delta \alpha _{4}(T)$, are strongly damped by viscosity. In $^{3}$He the
short-wave-length thermal ripplons are strongly damped in the whole
temperature range. Therefore, the ripplon contribution to $\Delta \alpha
_{4}(T)$ is small. The long-wave-length ripplons with energy $\hbar \omega
_{k}\ll T$ are not damped by viscosity, but their contribution to $\Delta
\alpha _{4}(T)$ is small because of the small phase volume.

The temperature dependence of the chemical potential $\mu _{3}\left(
T\right) $ in $^{3}$He, obtained using Eqs. (\ref{mu}),(\ref{Fv}) from the
experimental data in Ref. \cite{MuTExp}, is shown in Fig. \ref{FigMu3}. At $%
T>0.3K$, where $^{3}$He is nondegenerate, this dependence with very high
accuracy coincides with the theoretical prediction\cite{DyugJLTP90}
\begin{equation}
\mu _{3}\left( T\right) =\mu _{3}\left( 0\right) -T\ln 2+\varepsilon
_{3}^{0}-T^{2}/T_{3}^{0},  \label{mu3T}
\end{equation}%
where
\begin{equation}
\varepsilon _{3}^{0}=0.28K\text{ and }T_{3}^{0}=4.83K.  \label{eps3}
\end{equation}%
\ At $T<0.3K$, when $^{3}$He is a degenerate Fermi liquid,
\begin{equation}
\mu _{3}\left( T\right) \approx \mu _{3}\left( 0\right) -T^{2}/2T_{F},
\label{mu3Tl}
\end{equation}%
where $T_{F}=0.36K$. Note, that the quadratic temperature dependence of the
chemical potential [see Eqs. (\ref{mu4}),(\ref{mu3T})] is the characteristic
feature of all quantum nondegenerate liquids.\cite{DyugJLTP90,DyugSSR90}
Such temperature dependence $\mu \left( T\right) $ also takes place in
liquid hydrogen.\cite{WeJETPLett2008} An alternative theory of this
dependence is given in Ref.\cite{AndreevT2}.

Similarly to the case of $^{4}$He, in $^{3}$He$\ \Delta _{3}\left( T\right)
/T$ is almost independent of temperature with accuracy $\sim 5\%$ in the
wide temperature range $0.3K<T<2.5K$ (see insert in Fig. \ref{FigMu3}).
According to Eq. (\ref{OmFermion}), this occasional feature of the function $%
\Delta _{3}\left( T\right) /T$ results in the nearly quadratic dependence $%
\Delta \alpha _{3}(T)$ in the same temperature interval (see Fig. \ref%
{FigSig3}). From comparison with the experiment on $\Delta \alpha _{3}(T)$
in Ref. \cite{SurTensExp3} we obtain the values $\Delta _{3}\left( 0\right)
=0.25K$ and $M_{3}^{\ast }\approx 2.25M_{3}^{0}$.

One sees, that the account for the temperature dependence of the chemical
potential substantially improves the agreement between the proposed theory
and the experimental data on $\Delta \alpha (T)$ for both helium isotopes
and extends the temperature interval of this agreement to the whole interval
of the liquid phase (see Figs. \ref{Fig2} and \ref{FigSig3}). According to
the above calculations and to the experimental data, in a wide temperature
range the deviation of the surface tension coefficient $\Delta \alpha
_{3}(T)\propto T^{2}$ and $\Delta \alpha _{4}(T)\propto T^{2}$. This shows
that the particle statistics of surfons is not very important for the
surface phenomena on the liquid helium. The extracted parameters of the
surfon activation energy and their effective mass are in a reasonable
agreement with the theoretical predictions (\ref{DE4}),(\ref{DE3}),(\ref{m*n}%
) of our rough model and can be used for more elaborated theoretical study
of the structure of this new type of surface excitations.
\begin{figure}[tbh]
\includegraphics[width=0.49\textwidth]{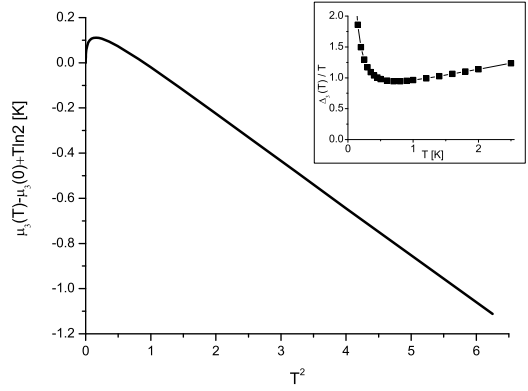}
\caption{The experimental temperature dependence of the chemical potential $%
\protect\mu _{3}(T)-\protect\mu _{3}(0)+T\ln 2$, obtained from the data in
Ref. \protect\cite{MuTExp} and plotted as function of $T^{2}$ to show that
Eq. (\protect\ref{mu3T}) is fulfilled with high accuracy. The insert figure
shows the temperature dependence of the surfon gap $\Delta _{3}\left(
T\right) $ divided by temperature. One sees that $\Delta _{3}\left( T\right)
/T$ is almost independent of $T$ in the wide temperature interval $%
0.3K<T\lesssim 2.5K$. This dependence, substituted to Eq. (\protect\ref%
{OmFermion}), explain the quadratic temperature dependence of the surface
tension $\Delta \protect\alpha _{3}(T)$ at $T>0.3K$.}
\label{FigMu3}
\end{figure}
\begin{figure}[tbh]
\includegraphics[width=0.49\textwidth]{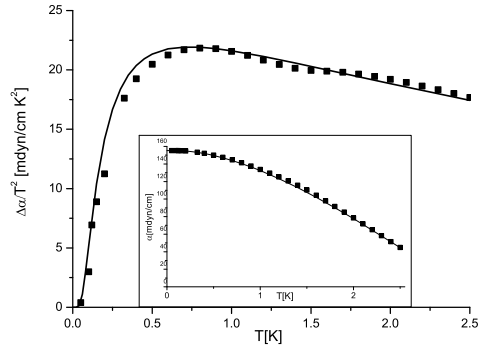}
\caption{The temperature dependence of the surface tension of $^{3}$He,
calculated using Eqs. (\protect\ref{OmFermion}),(\protect\ref{DeltaT3})
(solid line) compared to the experimental data from Ref. \protect\cite%
{SurTensExp3} (black squares). The agreement is so high that in the linear
scale (insert figure) the deviation cannot be detected by eye. Therefore, on
the main figure we plot $\Delta \protect\alpha _{3}/T^{2}$. The agreement
between theory and experiment considerably improves after taking the
experimental dependence of $\protect\mu \left( T\right) $ into account.}
\label{FigSig3}
\end{figure}

\section{Discussion and summary}

In Sec. II we proposed a simple theoretical model of the new type of surface
excitations (called surfons), and perform the quantum-mechanical calculation
of the surfon activation energy $E_{s}$ and their effective mass $M^{\ast }$
basing on this model. This calculation substantiates the existence of
surfons, because it shows that He atoms have at least one discrete energy
level at the liquid surface. He atoms are attracted to the surface by the
van der Waals force and become localized in the direction, perpendicular to
the surface, propagating only in the surface plane. As the result, they form
a 2D gas on the liquid surface. The temperature-dependent concentration of
the surfons is determined by their activation energy $\Delta =E_{s}-\mu $
[see Eq. (\ref{Delta})]. If $E_{s}<\mu ,$ the liquid is unstable. If $E_{s}>0
$, the concentration of surfons in negligibly small compared to the
concentration of helium vapor at any temperature. For both He isotopes, the
calculated values $E_{s}$ lie in the interval $\mu <E_{s}<0$. The formation
of a dimple under the surfon reduces the surfon activation energy by the
value $\Delta E_{s}\sim 1K$ [see Eqs. (\ref{DE4}),(\ref{DE3})]. The dimple
also increases several times the effective mass of the surfons in their
in-plane motion [see Eqs. (\ref{m*}),(\ref{m*n})]. This increase is not
surprising, because during the in-plane motion the surfon also drags the
dimple, which includes the motion of several atoms in the liquid.

The experimental observation and investigation of surfons is possible
because the surfons make the main contribution to the temperature dependence
of the surface tension. In Sec. III we calculate this contribution, taking
into account the temperature dependence of the chemical potential and liquid
density, which we derive from the experimental data. This considerably
improves the agreement between our theory and experiment of the temperature
dependence of the surface tension as compared to the previous letter\cite%
{SurStates}, making the deviations to be as small as $\lesssim 1\%$ in the
whole temperature range [for $^{4}$He this agreement is achieved after the
inclusion of ripplon contribution, given by Eq. (\ref{Atkins}) at
temperature below the $\lambda $-point]. The two extracted fitting
parameters, the surfon activation energy $\Delta $ and the effective mass $%
M^{\ast }$, are in a reasonable agreement with our calculations in Sec. II.
 Their values are given by Eqs. (\ref{D})-(\ref{M}). 
 At very low temperature the surfon contribution to the
temperature dependence of the surface tension is small,
because the concentration of surfons is exponentially small being determined
by the activation energy $\Delta $. At high temperature the surfon
contribution $\propto T^{2}$ for both helium isotopes.

For even higher agreement between theory and experiment on the temperature
dependence of the surface tension we also considered the interaction between
surfons. At low temperature this interaction is not important, because the
concentration of surfons is low. At higher temperature the interaction
between surfons only renormalizes the surfon effective mass. The details of
these results will be published elsewhere.

\section{Acknowledgement}

A.G. thanks Prof. A.F. Krutov for useful discussions. The work was supported
by the Foundation "Dynasty" and by the visitor program of Max Planck
Institute for the Physics of Complex Systems.


\begin{thebibliography}{99}
\bibitem{Comment1} This exchange energy correction is expected to be small
and less important for the value of the surface energy level than the local
adjustment of the atoms in the liquid to the surfon. The small value of the
exchange energy comes from the very small overlap of the wave functions of
He atoms due to their hard-core repulsion.
\end{thebibliography}
\end{document}